\documentstyle[12pt]{article}
\newcounter{mycount}

\newcommand{\be}{\begin{eqnarray}}
\newcommand{\ee}{\end{eqnarray}}
\newcommand{\bfl}{\begin{flushleft}}
\newcommand{\efl}{\end{flushleft}}
\newcommand\ie {{\it i.e. }}

\newcommand\half{\frac 1 2 }

\begin{document}

\centerline{\Large\bf A Coherent State  Path Integral for Anyons }
\vspace* {-35 mm}
\begin{flushright}  USITP-94-13 \\
September-1994
\end{flushright}
\vskip 55mm
\centerline{\bf J. Grundberg$^{\star}$ and T.H. Hansson$^{\dagger}$}
\vskip 15mm
\newcommand \sss {\mbox{ $<\overline{s}s>$} }
\def\fk{\mbox{ $f_K$} }
\centerline{\bf ABSTRACT}
\vskip 3mm
We derive an $su(1,1)$ coherent state path integral formula
for a system of two one-dimensional anyons
in a harmonic potential. By a change of variables we
transform this integral into a coherent states path
integral for a harmonic oscillator with a shifted energy.
The shift is the same as the one obtained
for anyons by other methods. We justify the procedure by showing that the
change of variables corresponds to a $su(1,1)$ version of the
Holstein-Primakoff transformation.

\vfil
\noindent
$^{\star}$ Department of Mathematics and Physics, M{\"a}lardalens
H{\"o}gskola, Box 11,\\S-72103 V{\"a}ster{\aa}s, Sweden\\
$^{\dagger}$Fysikum, University of Stockholm, Box 6730, S-11385 Stockholm,
Sweden\\
$^{\dagger}$Supported by the Swedish Natural Science Research Council

\eject

\bibliographystyle{nphys}

In this paper we construct a coherent state path integral for a pair of one
dimensional ''anyons''. It has often proved fruitful to have several methods
to study one and the same system, and this is our excuse for offering yet
another way of deriving a well-known result.

In one and two space dimensions, quantum mechanical wave functions for many
particle systems need not be single valued, since the corresponding
configuration space ${\cal M}$ is not simply connected. In two dimensions,
the phases associated
with closed curves in ${\cal M}$, corresponding to exchange of identical
particles, can be interpreted in terms of particle statistics, and depend on
a real parameter $\theta$\cite{lein1}. Under clockwise exchange of a single
pair
of particles the
wave function acquires the phase $e^{i\theta}$, so $\theta = 0 $ and $\theta =
\pi $ correspond to bosons and fermions respectively.

In one space dimension, the concept of oriented exchange is not meaningful,
but one can still define fractional statistics in several different ways.
Of particular interest is the  operator, or ''Heisenberg'', approach
to quantization, where the basic objects are
 unitary representations of the algebra of observables defined on
phase space. In the case of identical particles, operators like $\vec x$ and
$\vec p$ (here and in the following we only deal with relative coordinates and
momenta) are not observables since they change under permutations. For the
case of two particles in one dimension the basic observables must thus be
quadratic in $\vec x$ and $\vec p$ and a convenient choice is:
 $A = \frac 1 4 (a^\dagger a +  aa^\dagger)$,
$B_+ = \half(a^\dagger)^2$, and $B_- = \half a^2$, where $a=(x+ip)/\sqrt 2$.
These operators satisfy the $ su(1,1)$ (or $sp(1,1)$) algebra\cite{lein3,pere2}
\be
[A, B_\pm ] = \pm B_\pm \ \ \ \ \ \ \ \
[B_+,B_-] = -2A. \label{algebra}
\ee
The relevant irreducible representations, which are discrete,
infinitely dimensional, and bounded from below,
are characterized by a real parameter $\mu$ and satisfy
\be
A|k,\mu\rangle &=&
(\mu + k )\, |k, \mu \rangle \label{Ak} \\
B_+ |k,\mu\rangle &=& \sqrt{(k+1)(k+2\mu)} |k+1,\mu\rangle \\
B_- |k,\mu\rangle &=& \sqrt{k(k+2\mu -1 )} |k-1,\mu\rangle  \\
\Gamma|k,\mu\rangle &=& \mu(\mu-1)\, |k,\mu \rangle \ \ \ \ \ ,
\label{gamma}
\ee
where $\Gamma = A^2 -\half(B_+B_- + B_-B_+)$ is the
quadratic Casimir operator, $\mu > 0$ and $k=0,1,2..$. Different values
for the real parameter $\mu$ correspond to inequivalent representations,
and $\mu =1/4 \ {\rm mod(1)}$ for bosons and $\mu =3/4 \ {\rm mod(1)}$ for
fermions. For general $\mu$ this algebra describes particles with fractional
statistics, hereafter referred to as anyons. In fact, it has been shown
that these one-dimensional ''anyons'' can be interpreted as ordinary
two-dimensional anyons in a magnetic field and restricted to the lowest
Landau level, if the two and one dimensional statistics parameters $\theta$
and $\mu$ are related by $\nu \equiv\theta/\pi = 2\mu - 1/2$.

We  now derive a path-integral formula for the anyon propagator. The
normal Feynman construction is not possible since we do not have conjugate
coordinates and momenta. We will instead use the coherent state
approach.
We start by defining coherent states by\cite{pere2}
\be
|\zeta \rangle = (1-|\zeta|^2)^\mu e^{\zeta B_+} |0,\mu\rangle \label{coh}
 \ \ \ \ ,
\ee
where $B_-|0,\mu\rangle = 0$. For $\mu> 1/2 $ the
resolution of unity is given by
\be
1 = \frac {2\mu-1} \pi\int_{|\zeta |<1}
d^2\zeta\frac{|\zeta\rangle\langle\zeta|}{(1-|\zeta|^2)^2} .
\ee
and following the standard route we get the path integral
\be
&&K(\zeta_i,\zeta_f) = \langle \zeta_f|e^{-2iAT} |\zeta_i\rangle = \cr\cr
  &&\int_{|\zeta_{t}|<1}  \prod_{t} \left[ \frac {(2\mu-1)d^2\zeta_{t}}
         {\pi(1-|\zeta|^2)^2}\right] \exp\left[ \mu\int_{0}^{T} dt \left
( \frac {\dot{{\overline \zeta}}\zeta - {\overline \zeta} \dot{\zeta}}
{1-|\zeta |^{2}}
        -2i\frac {1+|\zeta |^{2}} {1-|\zeta|^{2}} \right) \right] ,
\ee
where we took $2A$ as our Hamiltonian, corresponding to a harmonic oscillator.
By a change of variables
\be
\beta = \sqrt{2\mu}\frac \zeta {\sqrt{1-|\zeta|^2}} ,
\label{change}
\ee
(6) simplifies  to
\be
K = \int \prod_t \left[ \frac {2\mu-1} {\pi \mu}d^2\beta_{t}\right]
  \exp\left\{ i \int_0^T dt\left [\frac{
     \dot{\overline \beta} \beta - \overline \beta \dot{\beta}}{2i}
   -i2(|\beta|^2 +\mu)\right] \right\}    \label{path} \ \ \ \ \ .
\ee
Comparing with the standard coherent state path-integral for the harmonic
oscillator\cite{schu1,jain5}
we see that (\ref{path}) corresponds to a harmonic oscillator
with energy eigenvalues
\be
E_{n} = 2n+2\mu = 2n + 1/2 + \nu \ \ \ \ \ . \label{spectrum}
\ee
This is indeed the correct result. For $\nu = 1$ we
recover the case of two fermions. Note that even though the original
Hamiltonian, $a^\dagger a + \frac 1 2 $, was that of an ordinary harmonic
oscillator,
the symmetry requirement removes every second level so the level spacing is
$2$ rather than $1$. Note further that (\ref{spectrum})
is correct also for bosons ($\nu = 0$ or $\mu = 1/4$)
even though the resolution of unity is valid only for $\mu > 1/2$.

That the spectrum is given by (\ref{spectrum}) is of course built in
by specifying the
property (2) of the representation.
The interesting thing about our
path integral derivation  is that it actually does give the right
result even though it at first sight looks rather dubious.
Recall that we made a change of variables in the naive version of the path
integral, {\em i.e.}, without properly defining it by an explicit
discretization,
and then interpreted the result as the naive form of the coherent states path
integral for the harmonic oscillator. To show how this can go wrong,
suppose we had interpreted (\ref{path}) as the naive
form of the phase space path
integral. For the harmonic oscillator, this
differs from the coherent states path integral
in that the zero-point energy is explicit
in the latter\cite{jain5}. Thus such an interpretation of (\ref{path})
would imply the incorrect spectrum
\be
E_{n} = 2n+1+\nu \ \ \ \ \ .
\ee

We now justify the path integral (\ref{path}) by showing that the
change of variables (\ref{change}) can be extended to a relation between
operators which directly relates the representation (2) - (5) to a
harmonic oscillator. This relation can be understood as
an $su(1,1)$ version of the
Holstein-Primakoff transformation\cite{hols1}.

We first note that the "classical" version of the generators of $su(1,1)$
is
\be
B^{\it cl}_- &=& \langle \zeta | B_- | \zeta \rangle  =  \frac {2\mu\zeta}
{1- |\zeta|^2 } \cr
B^{\it cl}_+ &=& \langle \zeta | B_+ | \zeta \rangle =
\frac {2\mu\overline\zeta} {1-|\zeta|^2} \cr
A^{\it cl} &=& \langle \zeta | A | \zeta \rangle = \mu
\frac{1+|\zeta|^2}{1-|\zeta|^2} \ \ \ \ \ .
\ee
In terms of $\beta$ we get
\be
B^{\it cl}_- &=& \beta (|\beta|^2 + 2\mu )^{1/2}\cr
B^{\it cl}_+ &=& \overline \beta (|\beta|^2 + 2\mu )^{1/2}\cr
A^{\it cl} &=& |\beta|^2 + \mu \ \ \ \ \ .\label{classical}
\ee
The Lagrangian in (10) is already on Hamiltonian form so we can directly
read off the Poisson bracket $\{\beta, \overline\beta\} = i$.
It is now easy to show that the Poisson bracket algebra of
$A^{\it cl}$, $B_+^{\it cl}$ and $B_-^{\it cl}$ is
$su(1,1)$ (taking into account the usual factor of $i$ between
commutators and Poisson brackets).
The operator version of (\ref{classical}) is
\be
\hat B_- &=& (b^\dagger b + 2\mu )^{1/2} b\cr
\hat B_+ &=& b^\dagger (b^\dagger b + 2\mu ) ^{1/2} \cr
\hat A &=& b^\dagger b + \mu  \ \ \ \ \ ,\label{opch}
\ee
where $[b,b^\dagger ] = 1$, and one again easily verifies, by direct
calculation, that $\hat B_-$, $\hat B_+$ and  $\hat A$ satisfies
the $su(1,1)$ algebra. Alternatively defining
$|k,\mu\rangle = (b^\dagger)^k/\sqrt{(k !)} |0, \mu\rangle$,
(15) implies (2) - (4).
Note that $b^\dagger$ creates two units of energy corresponding to the
level spacing appropriate for identical particles, while $a^\dagger$
creates one unit of energy appropriate for distinguishable particles.

Equation (\ref{opch}) is the promised $su(1,1)$
version of the Holstein-Primakoff transformation. If we now derive a path
integral using the coherent states $|\beta\rangle = e^{\zeta b^{\dagger}}
|0\rangle$ (where $b|0\rangle = 0$), and the Hamiltonian $2\hat A$
we arrive at
(\ref{path}) (except for a numerical factor in the measure). This is
clearly completely consistent with our interpretation of  (\ref{path})
as a harmonic oscillator coherent states path integral.

In previous work it has been shown that two anyons on a line is equivalent
to two bosons or two fermions interacting via a two-body potential of the
form $V(x) = \nu(\nu - 1)/(2x^2) $\cite{lein3,poly4}.
In the case of noninteracting
anyons Polychronachos made this equivalence explicit by constructing a
unitary transformation. It is natural to ask if we can recover the $1/x^2$
-potential from the path integral. We again make a change of variables in the
naive form of the path integral. Instead of $\overline\beta , \beta$ we
introduce the phase space coordinates $x$ and $p$ defined by
\be
x^2 &=& 2 A^{cl} + B^{cl}_+ + B^{cl}_-
         = |\beta + \sqrt{|\beta|^2 +2\mu}\thinspace|^2\cr
ixp &=& B^{cl}_{-} - B^{cl}_{+}
         = (\beta - \overline\beta )\sqrt{|\beta|^2 + 2\mu}\ \ \ \ \ .
\ee
If we require that $x$ and $p$ satisfy the canonical relation $\{x,p\}=1$
we can show, using the $su(1,1)$ Poisson bracket algebra,
that $p^2$ must be of the form
\be
p^2 = 2 A^{cl} -  B^{cl}_+ - B^{cl}_- + \frac{\lambda}{x^2} \ \ \ \ \  .
\ee
The constant  $\lambda$ is not fixed by the algebra, but by the
specific representation. Evaluating the Casimir invariant using its
definition in terms of $\hat A$, $\hat B_+$ and $\hat B_-$ we get,
using (13), (15) and (16),
 $\Gamma = \lambda/4 = \mu^2$, or $\lambda = (2\mu)^2$, which implies
\be
2A = \frac 1 2 p^2 +\frac{(2\mu)^2}{2x^2} + \frac 1 2 x^2 .
\ee
This construction is nothing but a phase space version of the derivation in
\cite{lein3}. We have indeed obtained a $1/x^2$ potential, but the
coefficient is $(2\mu)^2=(\nu+ 1/2)^2$  rather than $\nu(\nu - 1)$.
This can be understood since, as noticed in \cite{lein3}, after
quantization the relation $\lambda = 4\Gamma$ is replaced by
$\lambda_q = 4\Gamma_q + 3/4$ or $\lambda_q = 4\mu(\mu-1) +3/4 =
(2\mu - 1/2)(2\mu - 3/2) = \nu(\nu - 1)$.

A final comment is that the choice (\ref{coh}) of coherent states was crucial
in order to get a simple path-integral formula. An alternative
way to define coherent states is as eigenstates
of the lowering operator,  \ie,
$B_-|\alpha\rangle = \alpha |\alpha
\rangle$\cite{baru1}. These coherent states, which have recently been used to
discuss Berry phases in anyon systems\cite{hans4},
gives the path integral as
a  messy expression involving Bessel functions of the second kind.

\end{document}